# Use of real-time Fourier Transform Infrared Reflectivity as an in situ monitor of YBCO film growth and processing

Gertjan Koster[1], Jeong-Uk Huh[1], R.H. Hammond[1] and M.R. Beasley[1]

[1] *Geballe Laboratory for Advanced Materials, Stanford University, Stanford, California, 94305, United States of America.*

**Fourier Transform Infrared (FTIR) spectroscopy has been utilized during high rate E-beam evaporation/deposition of YBa$_2$Cu$_3$O$_7$ (YBCO). The results demonstrate the great utility of FTIR as an in situ monitor of YBCO deposition and processing. We detect different (amorphous/fine polycrystalline) insulating pre-existing phases to the high T$_c$ superconducting phase which appear to have distinct reflectivity fingerprints dominated by thin film interference effects, as a function of temperature and oxygen pressure. These fingerprints reveal some of the kinetic and thermodynamic pathways during the growth of YBCO.**

As the technology of complex oxide thin film growth progresses, so too does the need for effective *in situ* characterization and monitoring methods. This progress has been stimulated in part by the efforts over the past two decades in the field of high temperature superconductivity and its applications, but now extends to the growth of complex oxides in general. For example, standard tools have been adapted (e.g., high pressure RHEED [1]) to accommodate the special conditions required for oxide synthesis. In this paper we demonstrate the utility of Fourier Transform Infrared (FTIR) spectroscopy as a new member of the arsenal of tools effective in the deposition of high temperature superconducting thin films of YBCO. FTIR can be used to monitor the absolute temperature of a growing film and, perhaps more importantly, its dielectric optical properties. The later can be used to monitor transitions and transformations that arise both during the deposition of the films and later in any subsequent processing. We have used both of these possibilities in the case of YBCO. More specifically, here we will focus on one of the candidate processes for $YBa_2Cu_3O_7$ (YBCO) coated conductors, high-rate (>100 Å/sec) electron beam (E-beam) deposition [2,3]. However, we anticipate that such uses of FTIR will have broader applicability for complex oxide thin films generally. In this paper the FTIR process is discussed as a monitor of film growth and subsequent phase changes.

We monitor the deposition and growth of our YBCO films using real time FTIR spectroscopy in the so-called specular Reflectance-Radiance mode. In this mode, the sum of specular radiance and specular reflectance is measured followed by measurement of the radiance alone by means of chopping. From these data both the reflectance and a black body temperature of the film can be derived, the latter by using the reflectance to normalize for variations in emittance of the sample in real time (a similar approach was reported by Badano et al. [4]).

As we shall see, strong thin film interference fringes appear in the reflectance spectrum during deposition and upon subsequent growth due to oxygenation. The two types of substrates we investigate in this paper consist of single crystal $SrTiO_3$ coated with a layer of 3000 Å $SrRuO_3$ (Figure 1) and IBAD metal alloy tape (Figure 2 and 3). Details are described elsewhere in [5,6,7].

In Figure 1 we show an example of an absolute temperature measurement of one of our samples using the FTIR system as described above. The FITR derived temperature of the surface compared to the heater block temperature is different due to the existence of a thermal gradient across the substrate thickness and the substrate heater block interface. As can be seen from the band-averaged reflectance, there is a strong discontinuity just after deposition has stopped and the cool down started. This discontinuity is much more visible in experiments performed on metal tape due to the strong contrast in optical properties with the film material, see below. For now, assume that this discontinuity is linked to the increasing oxygen pressure after deposition and originates in the various phase transitions the deposit undergoes rapidly. Presumably, this is accompanied by the quick release of latent heat, giving rise to a substantial increase in temperature as detected by FTIR (Note that the thermocouple temperature is continuously decreasing at all times after deposition has stopped). This example shows that our setup gives real time Differential Thermal Analysis (DTA) information.

The following experiments are performed on metal tape substrates. Figure 2 shows the sequence of reflectance data taken with a 1 second interval during a typical evaporation of Y, Ba and Cu on IBAD tape followed by the introduction of molecular oxygen into the deposition chamber. Important changes in oxygen pressure during this experiment are depicted as well as changes in the temperature of the sample holder.

In Fig. 2 we see that initially (t=0) the reflectance is essentially that of the metallic tape. Then, during the deposition stage at an oxygen pressure of $5 \times 10^{-5}$ Torr (stage A, first 120 seconds), the reflectance develops a first minimum which moves to lower energies until a second minimum appears, as a function of deposition time. This phenomenon is clearly the result of thin film interference, and as the thickness of the deposit increases, the number of visible periods in the measured bandwidth increases as well, consistent with the thickness. The overall intensity tends to decrease with increasing thickness, which probably means the deposit is absorbing almost to the point where the interference fringes are no longer visible anymore at the final stages of deposition. During this stage we have not obtained reliable estimates for n and k [8].

Progressing in time, looking at Fig. 2, upon terminating the deposition and introducing additional oxygen up to a few mTorr, (stage B, at t > 144 s), very abruptly the amplitude becomes much stronger. Apparently, the material is transforming into a more transparent substance upon absorption of oxygen (the refractive index n is estimated to be 2 and the absorption coefficient k small ~-0.04 in most of this spectral range from fits using a simple recursive method for multilayers). Also, judging from the period of the fringes, the film tends to decrease in thickness compared to stage A, which has been confirmed by subsequent profilometry. Note that we have not been able to separate out the effect of a change of refractive index, see also above. Assuming that the film mass is preserved during this stage (no de-sorption takes place), it indicates a significant densification of the material.

Continuing to add more oxygen, in Fig. 2, we find the conditions to be within the known stability region for YBCO, and we see that the amplitude of the interference fringes becomes smaller (stage C, t > 170 s) whereas the period remains constant. We infer from these observations as well as from subsequent *ex situ* X-ray scattering (performed in real time in a controlled atmosphere XRD chamber) that crystalline YBCO starts to precipitate from the precursor. Optically, the material becomes more absorbing. From the phase stability diagram in [2], we see that YBCO should be in equilibrium with liquid $BaCu_2O_2$.

After this stage, the temperature is lowered in combination with a further increase of the oxygen pressure. In stage D (t > 470 s), YBCO is no more in equilibrium with liquid $BaCu_2O_2$ [2]. FTIR detects this phase by strong dampening of the interference fringes. Again, it appears that the period of the fringes remains constant (thickness and n constant). Finally, at low temperature and several 100 Torr of oxygen, the reflectance shows that the interference fringes completely dampen followed by an increased signature of metallic Drude-like behavior (i.e., conducting YBCO).

Now we turn to some experiments where some of the stages described above are amplified. In Fig. 3 a), the reflectance of a sample deposited at 830 °C and cooled down at the stage (B) when the oxygen pressure is only a few mTorr is shown. In this case, the interference fringes do not change significantly and upon inspection *ex situ*, the sample appears to be fully transparent and shiny. X-ray scattering reveals no major crystalline phases. These "glassy" samples tend to be relatively stable in air. Both of the above processes have been used to form so-called "precursors", which have been later annealed or reacted to form YBCO films.

When we compare the above reflectance sequences with the reflectance measured for a sample that contains only Ba and Cu, Fig. 3 b), we observe similar features; however, some important details are different. First, no abrupt decrease of the strong fringes is observed, when the pressure and temperature are inside the stability region of YBCO. This seems trivial since no Y is present to react to YBCO, but it confirms that the observed changes described above are related to the crystallization of YBCO. A more gradual decrease in fringe amplitude occurs when the temperature and pressure are above the projected melting line of $BaCu_2O_2$. No metallic behavior is observed upon further cooling. Interestingly, when the sample is cooled (at constant oxygen pressure, about 1 Torr.) after being in the melted region, the disappearance of fringes observed above when the oxygen in raised too high is not seen: the fringes persist till about 600 °C, and then decrease in amplitude by half. We infer that the liquid $BaCu_2O_2$ has super-cooled and then frozen without transforming into the two solids (CuO and $BaCuO_2$).

If the deposition temperature is kept much lower (300 °C or lower) during deposition (oxygen pressure 5 x$10^{-5}$ Torr) (A), clear intensity oscillations are visible throughout the deposition stage, see Figure 3 c). The abrupt onset of fringes after termination of the deposition and introduction of oxygen is not observed, simply because strong fringes are already present. Apparently, the densification (period) and the large amplitude are observed as deposited. This indicates that we already have relatively high-density material throughout the deposition at lower temperatures, and that apparently the "glassy" phase is a thermodynamic phase, based on the finding that at 830 °C the transformation oxygen pressure is 3 mTorr, and at 300 °C it is less than 5 x $10^{-5}$ Torr.

Although not shown in this paper, we have measured the growth kinetics in real time by X-ray scattering using an ambient controlled hot stage as a function of stoichiometry, of deposition sequence and of layer thickness [9] using the precursor films prepared by interrupting the growth process when at a particular stage, as described above. Furthermore, the nature of the liquid or glassy phase has been studied by XPS and XRD on quenched films at various stages which reveal the following: Quenching right after deposition and no further oxidation leads to powder-like films which show no crystalline phases, are very unstable in air [10]. The partially oxidized material tends to be much more stable in air, is transparent and copper is predominantly mono-valent [11].

In conclusion, our data demonstrates clearly the utility of *in situ* FTIR in thin film high $T_c$ superconducting oxide deposition. In a previous paper, we argued the presence of a liquid phase assisting along with an activated oxygen flux in our e-beam deposited YBCO in situ film growth to explain both the growth morphologies observed by TEM imaging as well as the dendrite shaped precipitates on our as-grown YBCO films [2]. However, as the use of the FTIR during this process showed that the YBCO probably did not actually form until the molecular oxygen pressure was raised into the region of stable YBCO in the $P(O_2)$ vs. Temperature diagram. This was later confirmed by doing XRD in a controlled atmosphere chamber on the amorphous precursors. Thus the FTIR was instrumental in revealing the actual fact that it was not in situ growth but actually a two-step process. Most importantly, the observed densification and passivation upon initial oxidation of the as-deposited amorphous precursor phase appears to be essential to the subsequent transformation and growth of the YBCO. When the fringes change amplitude we detect YBCO precipitation from the glass-liquid giving us the possibility to controlled the rate of precipitation by oxygen pressure and substrate temperature. Finally, we believe that crystal growth via liquid or glass precursors has potential for growing many oxide films where the phase diagram does not permit equilibrium between the vapor and the crystal phases.

We would like to acknowledge Peter Rosenthal from MKS inc. and Vladimir Mathias from LANL for their help in this work. This work was supported by the DoD through the MURI program on critical scientific challenges for coated conductors.

**Figure Captions**

**Figure 1** Reflectance (Band average) and Temperature as a function of time during and after deposition of YBCO on a $SrRuO_3$ buffered $SrTiO_3$ substrate. The open black square indicate the temperature of the heater block measured with a Thermocouple, solid red circles are the temperature determined from the normalized blackbody spectrum measured from the sample surface. The blue curve is band averaged values found for the reflectance centered at wave numbers?

**Figure 2** Sequence of reflectance (R) spectra (1000-5000 $cm^{-1}$) recorded during deposition and post anneal of YBCO on metal tape (#434, $T_{deposion}$=830°C). The five detected phases are marked A-E and a dashed line roughly indicates a phase transition; **A-** amorphous (0-144s), **B-** dense amorphous (glass, 144-170s), **C-** YBCO-6 + liquid $BaCuO_2$ (170-350s), **D-** YBCO-6 + decomposed $BaCuO_2$ (350-750s), **E-** metallic YBCO-7

**Figure 3** Sequence of reflectance (R) spectra (1000-5000 $cm^{-1}$) recorded during deposition and post anneal of YBCO on metal tape for a) YBCO deposited at 830 °C and 5x $10^{-5}$ Torr $O_2$ pressure during post anneal limited to a few mTorr while cooled to low temperature (#442). Only phases A and B are detected; b) deposition (830 °Ç and 5 x $10^{-5}$ Torr) and post anneal of nominal composition $BaCu2O_2$ (#450). Phases A and B are detected (with the only difference that Y is not present and thus no YBCO), however, above ~3 Torr a new phase (C') is detected, i.e., decomposition to two solids, CuO and $BaCuO_2$.; c)YBCO deposited at 5 x$10^{-5}$ Torr, and $T_{deposion}$=$T_{post-anneal}$=300 °C (#489).

Only one amorphous phase is detected (B') which resembles the B-phase above but probably differs slightly in density as detected by nano-indentation experiments [12];

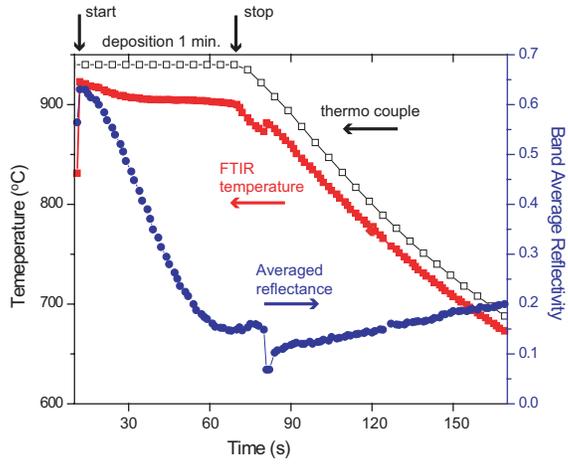

Figure 1 G. Koster et al.

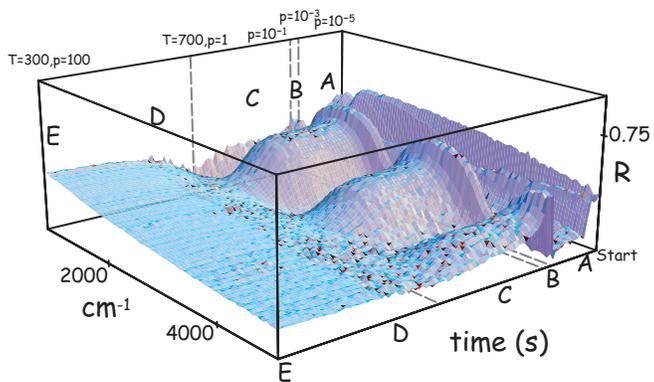

Figure 2 G. Koster et al.

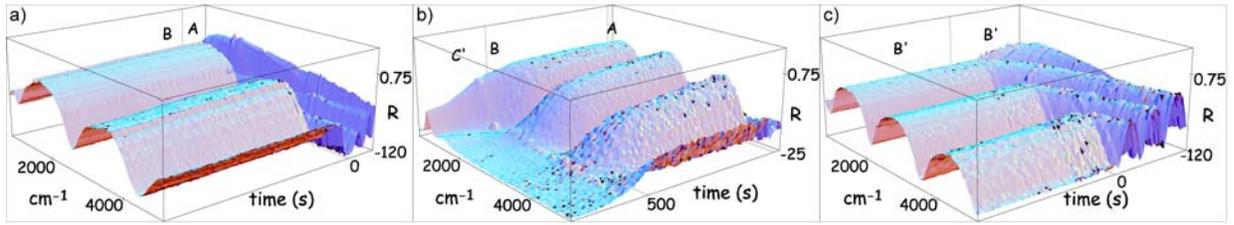

Figure 3 G. Koster et al